\documentstyle [12pt] {article}
\makeatother

\begin{document}
\baselineskip 20pt
\begin{titlepage}
\begin{center}
{\large\bf  STABILITY ESTIMATE IN SCATTERING}\\
{\large\bf THEORY  AND ITS APPLICATION TO}\\
{\large\bf MESOSCOPIC SYSTEMS AND}\\
{\large\bf QUANTUM CHAOS}\\ \ \\

{\bf Alexander G. Ramm $^{1}$ and Gennady P. Berman $^{2, 3}$ }\\ \ \\
$^1$Department of Mathematics,\\
Kansas State University, Manhattan, KS 66505-2602, U.S.A.\\
E-MAIL: RAMM@KSUVM.KSU.EDU\\ \ \\
$^2$Center for Nonlinear Studies, MS-B258\\
Los Alamos National Laboratory\\
Los Alamos, New Mexico 87545, U.S.A.\\
E-MAIL: GPB@GOSHAWK.LANL.GOV\\
and\\
$^3$Kirensky Institute of Physics;\\
Research and Educational Center for Nonlinear Processes\\
at The Krasnoyarsk Technical University;\\
Theoretical Department\\
at The Krasnoyarsk State University;\\
660036, Krasnoyarsk, Russia\\ \ \\ \ \\ \ \\ \ \\ \ \\
{\bf Abstract}
\end{center}
\quotation\noindent

We consider  scattering of a free quantum particle on a singular potential with
 rather arbitrary shape of the support of the potential.
  In the classical limit $\hbar=0$ this
 problem reduces to the well known problem of chaotic scattering. The universal
 estimates for the stability of the scattering amplitudes are derived. The
 application of the obtained results to the mesoscopic systems and quantum
chaos
 are discussed. We also discuss a possibility of experimental verification of
 the obtained results.
\endquotation
\end{titlepage}
{\bf I.  Introduction}\\ \ \\
Recently much attention has been paid to the theoretical and experimental
 investigations of the scattering of a free quantum particle on the obstacles
 with rather complicated form of boundaries. Of special interest are
 the studies of the scattering processes in mesoscopic systems at the
 ballistic regime when quantum effects and the geometry of the scattering
 potential are important [1-13]. Usually, these quantum systems are
 nonintegrable, and if they are treated classically they
  exhibit dynamical chaos, that is, strong
 (exponential) instability of motion under small variation
  of parameters (such
 as energy of an incident wave, form of the potential, etc).
 That is why
 one of the main problems in studying such systems is to
 determine the role and contribution of fluctuations and
 correlations in the scattering amplitudes and cross sections [14-22].

In this paper we consider a scattering problem for a free quantum particle
 scattered by a bounded obstacle with
 rather arbitrary shapes of the boundary. The boundary may consist
of several connected components.
Similar situation occurs in the processes of
 ballistic scattering in the mesoscopic systems widely considered nowadays
 [1-13]. The results obtained can be formulated in the following way. It is
 shown that there exists the region of parameters where small variation of
 rather arbitrary singular potential (note, that in this case the variation of
 the whole energy is infinite) leads only to small variations of the scattering
 amplitudes. This region of parameters can be defined as a region of strong
 correlations. These correlations are universal, and do not depend on the
 concrete structure of the resonances. We discuss the obtained results in
 connection with the general problem of quantum chaos and experimental
 observations of fluctuation and correlation effects in quantum chaotic
 scattering. The paper is
 organized as follows. In section 2 we present
  a stability estimate for the
 scattering amplitudes for rather wide classes of potentials. In
 section 3 a proof of the stability of the scattering
 amplitudes is given for  a singular
 potential. Applications to the quantum chaotic scattering
  are discussed in
 section 4.\\ \ \\
{\bf 2. Stability Estimate for the Scattering Amplitude}\\ \ \\
In this section we prove that small variations
 of the potential lead to small perturbations of
 the scattering amplitude for a class of strongly singular
 potentials which can take infinite values on sets of positive
  measure. The notion of small variations will be specified.

1. Let $D=\bigcup_{j=1}^JD_j$,
 $\Gamma:=\partial D=\bigcup_{j=1}^J\Gamma_j$,
 $D_j\subset R^n$ is a bounded domain with a
  $C^{2,\nu}$, $0<\nu\le 1$, boundary
 $\Gamma_j$. This means that in the local coordinates the equation of
 $\Gamma_j:=\partial D_j$ is $x_n=\phi(x^\prime)$,
 $x^\prime:=(x_1,x_2,...,x_{n-1})$, $\phi\in C^{2,\nu}$,
 $||\phi||_{C^{2,\nu}}\le\Phi_\nu$.

Assume $D\subset B_a:=\{x: |x|\le a\}$, and
  $D_j\bigcap D_i=\emptyset$ if
 $i\not=j$,
$J<\infty$. Define $u_0:=\exp(ik\alpha\cdot x)$. Define
$$
q(x;t):=t\chi_D(x),\quad \chi_D(x):=\cases{1, &in $D$,\cr 0, &in
 $D^\prime:=R^n\setminus D$,\cr}
$$
where parameter $t\in [1,\infty]$.
For definiteness take only $n=3$ in what follows. Consider the scattering
 problem
$$
[\nabla^2+k^2-q(x;t)]u=0\quad in\quad R^3.\eqno(1)
$$
$$
u=\exp(ik\alpha\cdot
 x)+A^{(t)}(\alpha^\prime,\alpha,k){{\exp(ikr)}\over{r}}+o\left({{1}\over{r}}
\right),\eqno(2)
$$
$$
 r:=|x|\rightarrow\infty,\quad {{x}\over{|x|}}:=\alpha^\prime.
$$
The scattering solution $u(x,\alpha,k;t):=u(t)$ is uniquely defined as the
 solution of (1), (2). It was proved in [26-28], that
$$
|u(t)-u_{\Gamma}|\rightarrow 0,\quad as\quad t\rightarrow +\infty,\eqno(3)
$$
where $u_\Gamma$ is the scattering solution to the obstacle scattering problem
$$
(\nabla^2+k^2)u_\Gamma=0\quad in\quad D^\prime,\quad u_\Gamma=0\quad
 on\quad\Gamma,\eqno(4)
$$
$$
u_\Gamma=u_0+A_\Gamma(\alpha^\prime,\alpha,k){{\exp(ikr)}\over{r}}+
o\left({{1}\over{r}}\right),\quad
 r=|x|\rightarrow\infty,\quad\alpha^\prime:={{x}\over{r}}.\eqno(5)
$$
The relation (3) has the following meaning
$$
||u(t)||_{L^2(D)}\le{{c}\over{\sqrt{t}}},\quad ||u(t)-u_\Gamma||_{H^2(\tilde
 D^\prime)}\le{{c}\over{t^{1/4}}},\eqno(6)
$$
$$
||u(t)||_{L^2(\Gamma)}\le{{c}\over{t^{1/4}}},\eqno(7)
$$
where $\tilde D^\prime$ is any compact strictly inner subdomain of $D^\prime$.
 Here and below $c>0$ denote various positive constants independent of $t$ or
 other parameters which vary.

Estimates (6),(7) are proved in [26-28]. It is proved in [24] that if $q_j(x)$,
 $j=1,2,$ generate the scattering amplitudes $A_j(\alpha^\prime,\alpha,k)$,
 then,
the following relation holds
$$
-4\pi A(\alpha^\prime,\alpha,k)=\int
 p(x)u_1(x,\alpha,k)u_2(x,-\alpha^\prime,k)dx,\eqno(8)
$$
where
$$
A:=A_1-A_2,\quad p:=q_1-q_2,\eqno(9)
$$
and $u_j$ is the scattering solution corresponding to $q_j$. Formula (8) is
 derived in [24] under the assumption that $q_j(x)\in L^p_{loc}$, $p>n/2$, and
 $q(x)$ is in $L^1(B^\prime_R)$, where $B^\prime_R:=R^3\setminus B_R$,
 $B_R:=\{x: |x|\le R\}$, $R>0$ is an arbitrary large fixed number.

In [29] an analog of (8) is derived for obstacle scattering. Namely, it is
 proved in [29] that if $\Gamma_j$, $j=1,2$,
 are bounded sufficiently smooth
 (say, Lipschitz) surfaces, and $A_j$ are the corresponding scattering
 amplitudes, $A_j:=A_{\Gamma_j}$, $A:=A_1-A_2$, then [29, formula (4)]
$$
-4\pi A(\alpha^\prime,\alpha,k)\eqno(10)
$$
$$
=\int_{\Gamma_{12}}[\bar
 u_{1N}(s,\alpha,k)u_2(s,-\alpha^\prime,k)-u_1(s,\alpha,k)u_{2N}(s,-\alpha^
\prime,k)]ds,
$$
where $N$ is the exterior unit normal to $\Gamma_{12}=\partial D_{12}$, where
 $D_{12}:=D_1\bigcup D_2$.\\
2. We claim that, uniformly in $t_j\in [1,\infty]$, the following
 stability estimate holds
$$
\sup_{\alpha^\prime,\alpha\in S^2; 0<k_1\le k\le k_2<\infty}
|A^{(t_1)}_{D_1}(\alpha^\prime,\alpha,k)-A^{(t_2)}_{D_2}(\alpha^\prime,\alpha,k)
 |\eqno(11)
$$
$$
\le c\{[min(t_1,t_2)]^{-1/4}+\rho(D_1,D_2)\},
$$
where $c=const.>0$, $c$ is independent on $t_j\in[1,\infty]$, and on
$D_j\subset
B_a$, $j=1,2$, such that $||\phi_j||_{C^{2,\nu}}\le\Phi_{\nu} $.

The distance $\rho(D_1,D_2)$ in (11)
is defined by the formula
$$
\rho(D_1,D_2):=
\sup_{x\in\partial D_1}\quad \inf_{y\in\partial D_2}|x-y|
$$
3. Note, that if $t\in [1,t_0]$, where $1<t_0<\infty$ is any fixed number, then
 the following estimate can be derived from (8)
$$
\sup_{\alpha^\prime,\alpha\in S^2; 0<k_1\le k\le
 k_2<\infty}|A^{(t_1)}_{D_1}(\alpha^\prime,\alpha,k)-A^{(t_2)}_{D_2}(\alpha^
\prime,\alpha,k)|\
$$
$$
\le {{c}\over{4\pi}}|t_1-t_2|{\int}_{D_1\bigcap
 D_2}dx+{{ct_0}\over{4\pi}}\int_{D_{12}\setminus(D_1\bigcap D_2)}dx
$$
$$
\le{{c}\over{4\pi}}|t_1-t_2||D_1\bigcap D_2|+{{ct_0}\over{4\pi}}\{|\partial
 D_1|+|\partial D_2|\}\rho(D_1,D_2)
$$
$$
\le c\{|t_1-t_2|+\rho(D_1,D_2)\}.\eqno(12)
$$
Here we have used the known estimate [24], [25]
$$
\max_{x\in R^3;\alpha\in S^2;0<k_1\le k\le k_2<\infty}|u_j|\le c.\eqno(13)
$$
In (12) $|\partial D_j|$ denotes the area of the surface $\partial D_j$, and
 $|D_1\bigcap D_2|$ denotes the volume of $D_1\bigcap D_2$.\\
4. If $t_1=t_2=+\infty$, then, the stability estimate
$$
\sup_{\alpha^\prime,\alpha\in S^2;0<k_1\le k\le
 k_2<\infty}|A_1(\alpha^\prime,\alpha,k)-A_2(\alpha^\prime,\alpha,k)|\le
 c\rho(D_1,D_2)\eqno(14)
$$
follows from formula (10), since
$$
\sup_{s\in\Gamma_j;  \alpha\in S^2;
 0<k_1 \le k\le k_2< \infty}|u_{jN}(s,\alpha,k)|\le c,
$$
$$
\sup_{\alpha^\prime\in S^2; s\in \Gamma_{j+1}; 0<k_1 \le k\le k_2
 <\infty}|u_{j}(s,-\alpha^\prime,k)|\le c\rho(D_1,D_2).
$$
Here $\Gamma_3:=\Gamma_1$, $j=1,2$.

The basic result (11), which contains both stability estimates (12) and (14),
is
 of interest because the inequality (11) holds $\it uniformly$ $\it in$ $t$,
 $t\in[1,\infty]$.\\
5. As an example, we present here the results on the dependence $c(k)$
 in (14) for
 the special case of the scattering potential.
We claim that the constant $c$ in (14) is of the order $O(k^2)$ as $k$ goes to
 infinity,
under the following assumptions: i) $J=1$,
 ii) $s\cdot N>b>0$ for $s$ in $S_1$ ($S_1:=\Gamma$) and
for $s$ in the perturbed surface, say $S_2$; here $N$ is the outer normal to
 $S_1$ (or
$S_2$)  at the point $s, b>0$ is a constant independent of $s$, $k$ and other
parameters.\\
Proof of the claim: If ii) holds, then from the estimate (6)
 in [23,p.66] it follows that $||v||_{B_{R}}<c$, $c$ is always assumed to be
 independent
 of $k$, $v:=u-u_0$,
where $u$ is the scattering solution corresponding to $S_1$, and $u_0$
 is the plane wave.
 From this and the Helmholtz equation one gets $||v||_2<ck^2$, where $||v||_2$
is the Sobolev space $H^2$ norm. Let $|v_N|$ stay for the $L^2(S_1)$ norm of
$v_N$ on
$S_1$. Then, an interpolation inequality yields the desired estimate:
$|v_N|<ck^{3/2}$. This estimate implies the claim that the constant $c$ in (14)
is of the order  $O(k^2)$ as $k$ grows to infinity. Indeed, estimating
integrals
 in (10)
by Cauchy's inequality one gets the sum of the products of the terms
of the type $|v_N|$ $|v|$ and terms of lower order in $k$ which are
easy to estimate by $O(k^{3/2})$. By an interpolation inequality, the norm
$|v|$ is $O(k^{1/2})$, so the result follows.
Let us formulate the known interpolation inequalities used above
(see [28])
$$
||D^rv||_{L^2(S_1)}<c{ t^{3/2-r}||v||_2 +t^{-1/2-r}||v||},\eqno(15)
$$
where $||v||$ is the $L^2$ norm in $B_a\setminus D, \partial D =S_1 $,
 $t>0$ in (15) is an  arbitrary
parameter, and $r=0$ or $1$.
 Take $r=0$ in (15) and minimize in $t>0$ the right-hand
side of (15), using the formulas $||v||_2<ck^2$, $||v||<c$, to get
for the right-hand side the estimate $O(k^{1/2})$ .
Similar argument for $r=1$ yields the
estimate $O(k^{3/2})$ as claimed.\\
{\underbar {Remark.}} The order in $k$ as $k \rightarrow \infty$
  in the estimate for the constant  $c$ in
(14), is not optimal. The optimal order is probably $O(1)$. For a ball,
for instance, we can prove that $|v_N|=O(k)$, rather than $ O(k^{3/2})$ and
$|v|=O(1)$, rather than $O(k^{1/2})$. This yields $c=O(k)$ as
$k \rightarrow \infty$.
The estimate based on the Cauchy inequality, used in our derivation, does
not take into account possible cancellations during integration in (10) due
to oscillations of the integrand for large $k$. The optimal orders are:
 1) $O(1)$ for $|v|$, 2) $O(k)$ for $|v_N|$ , and 3) $O(1)$ for the
cross section as $k \to \infty$. These conclusions can be also obtained
from the geometrical optics approximation
 (see formula (150.16) in H.Honl,A.Maue, K.Westpfahl, Theorie
der Beugung, Springer Verlag, Berlin, 1961).\\
6. Let us formulate the result proved in [28].\\
{\underbar {Theorem 1}}. Under the assumption made in section 2.1,  estimate
 (11) holds with the constant $c>0$ independent of $t$, where
 $t\in[1,\infty]$, $D_j\subset B_a$, $\partial D_j\subset C^{2,\nu}$,
 $||\phi_j||_{C^{2,\nu}}\le\Phi_\nu$.

In section 3 the proof of estimate (14)
 is given for the case $t_1=t_2=\infty$ which is of interest in
 applications.  In section 4 applications are
 discussed.\\ \ \\
{\bf 3. Proof of the Stability Estimate (14)}\\ \ \\
Let us assume that
$$
q_j(x)=\cases{+\infty, &in $D_j$,\cr 0, &in $D_j^\prime:=R^n\setminus D_j, n\le
 2$.\cr}\eqno(16)
$$
This is the case discussed in section 2.4 (see formula (14)). We assume $n=3$
 for definiteness.  The argument is the same for $n\ge 1$.

There are three ways to prove estimate (14) under the assumption (16). One way
 is to take $t_1=t_2=+\infty$ in (11), and note that the
 right-hand side equals $c\rho(D_1,D_2)$ if $t_1=t_2=+\infty$.
  The second way, is to take
$t_1=t_2=t<\infty$, and then let $t\rightarrow +\infty$, and use
formula (8)  and estimates (6), (7).  These estimates allow one to
 derive formula (10) from which estimate (14) follows. Estimate (14) is a
 particular form of estimate (11) for the case  when
 $\min(t_1,t_2)=+\infty$. The third way is based on estimate (10). Let us
 use this way. We assume that the distance $\rho(D_j,D_m)$, $j\not= m$, is much
 greater than the distance $\rho(D_j,\tilde D_j)$, where $\tilde D_j$ is the
 perturbed domain $D_j$. The number $J$ of the connected components of the
 domain $D$ is fixed and finite. Therefore, the input of the variation of
 $\partial D$ in the scattering amplitude is of the order of magnitude of the
 input of the variation of $\partial D_j$ , $1\le j\le J$. Therefore,
 one may use formula (10) assuming that $\partial D$ has one connected
component
 $\partial D_1$, and $\partial D_2:=\partial \tilde D_1$
 is a small variation of
$\partial D_1$ in the sense that $\rho(D_1,D_2)$ is small. It follows from (10)
 that
$$
|A(\alpha^\prime,\alpha,k)|\le{{1}\over{4\pi}}\int_{\Gamma^\prime_1}|u_{1N}(s,
\alpha,k)u_2(s,-\alpha^\prime,k)|ds\eqno(17)
$$
$$
+\int_{\Gamma^\prime_2}|u_{1}(s,\alpha,k)u_{2N}(s,-\alpha^\prime,k)|ds:=I_1+I_2,
$$
where $\Gamma^\prime_1$ is the part of $\Gamma_1$ which lies outside $D_2$, and
 $\Gamma^\prime_2$ is the part of $\Gamma_2$ which lies outside $D_1$.

One can use the following estimates
$$
\gamma:=\max_{j=1,2}\sup_{s\in\Gamma_j; \beta\in S^2; 0<k_1\le k\le
 k_2<\infty}|u_{jN}(s,\beta,k)|\le c,\eqno(18)
$$
$$
\max_{j=1,2}\sup_{s\in\Gamma^\prime_j; \beta\in S^2; 0<k_1\le k\le
 k_2<\infty}|u_{j+1}(s,\beta,k)|\le c\rho(D_1,D_2),\quad u_3:=u_1,\eqno(19)
$$
and formula (17), to get the desired estimate (14). Let us discuss estimates
 (18) and (19).  The constant $c$ in (18) and (19) depends on the parameters
$k_1$, $k_2$, $a$, and on the parameter $\Phi_\nu$, which is introduced in
 section 2.1, and which describes the smoothness of the boundary:
 $||\phi_j||_{C^{2,\nu}}\le \Phi_\nu$. This constant does not depend on the
 particular choice of $D_j$. Let us prove the last claim. Suppose on the
 contrary,
that there exists a sequence $D_{jn}$ of the obstacles $D_{jn}\subset B_a$,
 $||\phi_{jn}||_{C^{2,\nu}}\le \Phi_\nu$, such that $\gamma_n\ge c_n$,
 $c_n\rightarrow\infty$, where $c_n$ are the constants in (18), (19),
 and $\gamma _n$ is $\gamma$ for the obstacle $D_{jn}$,
  $n=1,2,...\quad$. By the Arzela-Ascoli compactness theorem one can assume
that
$$
\phi_{jn}\stackrel{C^{2,\nu^\prime}}\rightarrow\psi_j,\quad
 0<\nu^\prime<\nu,\quad u_{jn}\stackrel{H^2_{loc}}\rightarrow u_j,\quad
 n\rightarrow\infty,
$$
where $u_j$ is the scattering solution corresponding to the limiting
 configuration of the surfaces $\Gamma_1$, $\Gamma_2$. For fixed surfaces
 $\Gamma_1$ and $\Gamma_2$, estimates (18) and (19) hold [23].

Note that it is sufficient to prove estimate (18). Indeed,
$$
|u_1(s,\beta,k)|=|u_1(s,\beta,k)-u_1(\tilde
 s,\beta,k)|\le\sup|u_{1N}(s,\beta,k)||s-\tilde s|
$$
$$
\le c\rho(D_1,D_2),
$$
where $s\in\Gamma^\prime_2$, $\tilde s\in\Gamma_1$,  $u_1(\tilde
 s,\beta,k)=0$, and the segment $\tilde ss$ is directed along the normal to
 $\Gamma^\prime_2$. A similar argument is valid for  $u_2(s,\beta,k)=0$,
 $s\in\Gamma^\prime_1$.

If $\Gamma_{jn}\rightarrow\Gamma_j$ in the sense
 $\phi_{jn}\stackrel{C^{2,\nu^\prime}}\rightarrow\psi_j$ as
 $n\rightarrow\infty$), then
$u_{jNn}\rightarrow u_{jN}$ as $n\rightarrow\infty$ (uniformly in
$s\in\Gamma_j$
 and in the parameters $\beta\in S^2$, $k\in[k_1,k_2]$, $0<k_1<k_2<\infty$),
 so that $\gamma_n\rightarrow\gamma$ as $n\rightarrow\infty$. Here $\gamma$ is
 the number defined by the left-hand side of (18) with $u_j$ corresponding to
 the limiting surfaces $\Gamma_j$. Since this $\gamma<\infty$, one obtains a
 contradiction: the inequality $\gamma_n\ge c_n\rightarrow+\infty$ contradicts
 to the equation $\gamma_n\rightarrow\gamma<\infty$. This contradiction proves
 that the constant $c$ in (18) and (19) does not depend on the particular
choice
 of the obstacles $D_j$ as long as the two conditions are satisfied:
$D_j\subset
 B_a$, $||\phi_j||_{C^{2,\nu}}\le\Phi_\nu$, and the parameters $a$, $\Phi_\nu$,
 $k_1$ and $k_2$ define the value of $c$ in (18), (19) and (14).\\ \ \\
{\bf 4. Appications to the Chaotic Scattering}\\ \ \\
The results derived above have direct application to the so-called chaotic
 scattering [14-22,30-32].
The problem of fluctuations of the scattering amplitudes and cross sections in
 the processes of elastic (and inelastic) collisions is well known, and has a
 long history (see [33-38] and references therein). In the elastic scattering
 which was considered in sections 1-3, these fluctuations of the scattering
 amplitudes can appear because of a high sensitivity to the details of the
 scattering: the parameters of the incident wave and the geometry of the
scatter
 potential. At the same time, the coherent effects (correlations) are also
 present in the scattering processes in some region of parameters
 [21,22,33,34,39]. Thus, the problem arises: how does one separate and describe
 the random and the coherent effects in the scatttering processes, and how
does one measure their contribution in experiments?

The first theoretical investigations of the statistical properties
 (fluctuations) of the scattering amplitudes and cross sections were presented
 in [33-38] (Ericson fluctuations). According to [33,34], the main reasons why
 the scattering amplitudes become random are the following. Let an incident
wave
 (the first term in (2)) have a wave-length $\lambda=2\pi/k$ much smaller than
 the characteristic dimension $L$ of the region $D$ where the scattering
 potential $q(x)$ in (1) is located: $kL\gg 1$. Before escaping from the region
 $D$, the incident wave can be reflected a large number of times from the
 boundaries $\Gamma_j$ of the support of the potential $q(x)$.
  In this case, a wave close to a
 standing wave appears in the system. These ``quasi-standing" (or
 quasy-stationary) waves can be associated
 with the resonances in the scattering amplitude. Each $n$-th resonance is
 characterized by two main parameters: the energy $E_n$, and the
 width $\Gamma_n$ [40]. There is usually one more important parameter which
characterizes the spacing between the neighboring resonances: $\Delta E_n$.
 Because the process of scattering is completely defined, the scattering
 amplitudes should be reproducible in different experiments, provided
that all
 conditions remain identical.
However, as was mentioned above, under the condition $kL\gg 1$ the number of
 reflections of the incident wave in the region $D$ can be very large (in
 [33,34] also the following
inequality is assumed to be satisfied: $\Gamma_n/\Delta E_n\gg 1$, which is
 called the regime of overlapping levels). Then, a small variation of
parameters
 can completely change the ``trajectory" of the wave, and consequently, the
 phase of the scattering amplitude. These ideas were developed in [33,34,36] on
 the basis of the statistical approach [41].

Recently, the problem of fluctuations of the scattering amplitudes has
attracted
 additional interest in connection with the so-called ``chaotic (irregular)
 scattering" (CS) in chemical reactions, particle physics, mesoscopic systems
 and other areas of physics [14-22,30-32]. The investigations of the CS can be
conventionally
 divided into three groups: (1) classical CS (CCS), (2) semiclassical CS
 (SCS), (3) quantum CS (QCS). The basic ideas are associated with the CCS,
since
 only in this case the dynamical chaos occurs. The investigations of the
 CCS were stimulated  by the significant progress achieved recently in studying
of
the
  dynamical chaos in the classical bounded Hamiltonian systems [42-45]. The
 classical phase space in this case can be very complicated, and each of
the trajectories belongs to one of
 the following three classes: (a) stable periodical
 trajectories, (b) unstable periodical trajectories, (c) chaotic (unperiodical)
 trajectories. Dynamical chaos in bounded systems is stationary in the sense
that it does not disappear at large times ($T\rightarrow\infty$). The systems
 where the CCS takes place are unbounded, and the additional trajectories
 appear: (d) unbounded trajectories. In the case of  a singular potential
$q(x)$
 considered above the trajectories (a) can be absent (see, for example,
[15]), and the trajectories (b) and (c) represent a  ``repeller" $\Omega_R$
 [15]. For the trajectories (d) this repeller leads to the ``transient chaos"
 which was previously investigated in various bounded conservative and
 dissipative systems (see, for example, [46-48]).

The main achievements in the CCS are associated with the understanding of the
 following facts: (1) although, in the CCS a direct contribution in the cross
 section is connected with the trajectories (d), the influence of the
repeller $\Omega_R$ - bounded (trapped) trajectories  on the process of
 scattering and fluctuations plays a very important role; (2) the CCS is a
general
 phenomenon rather then an exception. (In some special cases of a singular
 potential [49], the set $\Omega_R$ can consist of only one unstable periodic
 trajectory). Usually, for singular potentials considered above, a repeller
 $\Omega_R$ is a Cantor set with a fractal structure (see, for example, paper
 [15] where an elastic scattering on three hard discs (3HD) was considered),
and
 is characterized by several quantities, such as the Hausdorff dimension $D_H$,
 Lyapunov exponents $\lambda_i$, the Kolmogorov-Sinai entropy per unit time
 $h_{KS}$, the escape rate $\gamma$, and other quantities (see [15] and
references
 therein). There are some relations between these parameters,
   for example, (see [15]):
$$
\gamma=\sum_{\lambda_i>0}\lambda_i-h_{KS}.\eqno(23)
$$
The escape rate $\gamma$ is a classical equivalent of the resonance width
 $\Gamma$: $\gamma\sim\Gamma/\hbar$ [15]. So, the relation (23) shows a
 fundamental property of the CCS: when a repeller $\Omega_R$ is chaotic
 ($h_{KS}>0$), the escape rate (and the resonance width $\Gamma$) is
decreasing.
 Also, in this case  large fluctuations appear in the quantities which
 characterize the process of CCS, for example, in the time delay function
 [15,20,22].

When one investigates the SCS and the QCS, the main problem is: what are the
 ``fingerprints" of the classical chaos on the quantum scattering ? For the
 first time, the problem of QCS was considered in [14], where the elastic
 scattering was studied on a two-dimensional surface of a constant negative
 curvature.  According to [14], the scattering phase shift as a function of the
 momentum is given by the phase angle of the Riemann's zeta function, and
 displays a very complicated (chaotic) behavior (see for details [14,21,22]).
In
 [16] the SCS was studied in the system of 3HD using the analysis based on the
 Gutzwiller trace formula [50]. This trace formula is valid when all periodic
 orbits of the repeller $\Omega_R$ are unstable and isolated. Both these
 conditions can be satisfied for the singular potential
 $q(x,t\rightarrow\infty)$
 considered in sections 1-3, including a particular case of a singular
 potential of the 3HD system considered in [15-18].

The quantum analysis presented in [21,22] shows that in the QCS the statistical
 properties of the fluctuations in the cross section can be described by the
 theory of random matrix ensembles [41]. Different aspects on the problem of
 fluctuations in the SCS and QCS are discussed in [14,16-19,21-39].

At the same time, much less is known about the contribution and characterisic
 properties of the correlations (coherent component) in the chaotic scattering.
 As  was pointed out in [33,34], a significant level of correlations in the
 cross section should be expected when, for example, the energy change $\delta
 E$ of the incident wave in (2) is small compared with the resonance width
 $\Gamma$
($\Gamma/\delta E>1$). According to [33,34], in this case essentially the same
 states are exited, and the scattering amplitudes are changed insignificantly.
 The existence of correlations in the QCS was discussed also in [21,22] for
some
 quasi-1D periodical potential (in [22] also an experiment is discussed in
 connection with the correlations in the chaotic scattering). It was shown in
 [21,22] that the correlations in energy for the matrix elements of the
 $S$-matrix exist, and exhibit themselves when $\Gamma/\delta E>1$, in
agreement
 with the Ericson hypothesis [33,34].

In connection with the problem of correlation effects in the quantum chaotic
 scattering, the consideration presented in sections 1-3 are of considerable
 interest. In particular, the estimate for the scattering amplitudes given by
 formula (14) is valid for the
  general case of singular potentials $q(x)$ supported in a compact
 region $D$. In this case the corresponding classical repeller
 $\Omega_R$ is generally chaotic. So, the result (14) means that even for
 classically chaotic (irregular) scattering, the strong quantum correlations in
 the scattering amplitudes exist in some region of parameters, and are of the
universal nature. The latter means that the quantum correlations in this region
 of parameters do not depend on the specific character of the resonance
 structure. The estimate (14) includes the constant $c$ which actually depends
 on the system's parameters
$$
c=c(k_1,k_2,a, \Phi_\nu)\eqno(24)
$$
That is why it is difficult to establish a relation between the region of
 parameters where the estimate (14) is valid, and the one  ($\delta
 E>\Gamma>\Delta E$) where the above discussed Ericson fluctuations are
 important.

The analytical and experimental investigations of the dependence (24) represent
 a significant interest for the further development of our understanding of the
 correlation effects in the processes of quantum chaotic scattering.

One of the possibilities to investigate the correlation and fluctuation effects
 in quantum chaotic scattering can be realized in the microwave experiments
 (see, for example, [51]). The main idea, which is used
  in these experiments, is
 that the Schr\"odinger equation for a free particle
  reduces to the Helmholtz
 equation which describes the propagation of the  classical waves. This
 correspondence was utilized in [51] to
 investigate the role of fluctuations in
 the chaotic scattering. In our opinion,
  this method is rather promising: it allows one to
 imitate the ballistic regime taking into account scattering,
  and to study the
 correlation effects in mesoscopic systems using a microwave
 technique.\\ \ \\
{\bf Acknowledgments}\\ \ \\
 AGR thanks NSF  for support. GPB  thanks
 Don Cohen, Gary Doolen and J.Mac Hyman of The Center
  for Nonlinear Studies, Los
 Alamos National Laboratory, for their hospitality.
\vfil\eject

{\bf References}\\ \ \\
1. Y. Imry, Europhys. Lett., 1 (1986) 249.\\ \ \\
2. G. Timp, A.M. Chang, P. Mankiewich, R. Behringer, J.E. Cunningham,
T.Y.Chang,
 R.E. Howard, Phys. Rev. Lett., 59 (1987) 732.\\ \ \\
3. G. Timp, H.U. Baranger, P.de Vegvar, J.E. Cunningham, R.E. Howard, R.
 Behringer, P.M. Mankiewich, Phys. Rev. Lett., 60 (1988) 2081.\\ \ \\
4. C.W.J. Beenakker, H. van Houten, Phys. Rev. Lett., 60 (1988) 2406.\\ \ \\
5. C.J.B. Ford, S. Washburn, M. B\"uttiker, C.M. Knoedler, J.M. Hong, Phys.
Rev.
 Lett., 62 (1989) 2724.\\ \ \\
6. H.U. Baranger, A.D. Stone, Phys. Rev. Lett., 63 (1989) 414.\\ \ \\
7. A.M. Chang, T.Y. Chang, H.U. Baranger, Phys. Rev. Lett., 63 (1989) 996.\\ \
 \\
8. H. Fukuyama, T. Ando (Eds.), Transport Phenomena in Mesoscopic Systems,
 Springer-Verlag, 1992.\\ \ \\
9. Physics of Low-Dimensional Semiconductor Structures, Edited by P.Butcher,
 N.H. March, M.P. Tosi, Plenum Publishing Corporation, 1993.\\ \ \\
10. Nanotechnology. Research and Perspectives, Edited by B.C. Crandall, J.
 Lewis, The MIT Press Cambridge, Massachusetts, London, 1992.\\ \ \\
11. Physics of Nanostructures, Edited by J.H. Davies, A.R. Long, Proceedings of
 the Thirty-Eighth Scottsh Universities Summer School in Physics, St Andrews,
 1991.\\ \ \\
12. G. Bauer, F. Kuchar, H. Heinrich (Eds.), Low-Dimensional Electronic
Systems.
New Concepts, Springer-Verlag, 1992.\\ \ \\
13. Semiconductors and Semimetals. Nanostructured Systems, Edited by M. Reed,
 Academic Press, Inc., 1992.\\ \ \\
14. M.C. Gutzwiller, Physica D, 7 (1983) 341.\\ \ \\
15. P. Gaspard, S.A. Rice, J. Chem. Phys., 90 (1989) 2225.\\ \ \\
16. P. Gaspard, S.A. Rice, J. Chem. Phys., 90 (1989) 2242.\\ \ \\
17. P. Gaspard, S.A. Rice, J. Chem. Phys., 90 (1989) 2255.\\ \ \\
18. P. Gaspard, Proceedings of the International School of Physics ``Enrico
 Fermi", Quantum Chaos,  North-Holland, 1993, p. 307.\\ \ \\
19. R. Bl\"umel, U. Smilansky, Phys. Rev. Lett., 60 (1988) 477.\\ \ \\
20. G. Troll, U. Smilansky, Physica D, 35 (1989) 34.\\ \ \\
21. R. Bl\"umel, U. Smilansky, Physica D, 36 (1989) 111.\\ \ \\
22. U. Smilansky, The Classical and Quantum Theory of Chaotic Scattering, in:
Chaos and Quantum Physics, Les Houches, North-Holland, 1991, p. 370.\\ \ \\
23. A.G. Ramm, Scattering by Obstacles, Reidel, Dordrecht, 1986.\\ \ \\
24. A.G. Ramm, Multidimensional Inverse Scattering Problems, Longman/Wiley, New
 York, 1992.\\ \ \\
25. A.G. Ramm, Acta Appl. Math., 28, N1, (1992) 1.\\ \ \\
26. A.G. Ramm, Izvestiya Vuzov, Mathematics, N5 (1965) 124; Math. Rev., 32
 $\#$7993.\\ \ \\
27. A.G. Ramm, J. Math. Anal. Appl., 84 (1981) 256.\\ \ \\
28. A.G. Ramm, Stability Estimates for Obstacle Scattering, 1993.\\ \ \\
29. A.G. Ramm, Appl. Math. Lett., 6,N5, (1993) 85.\\ \ \\
30. B. Eckhardt, Phys. Reports, 163 (1988) 205.\\ \ \\
31. B. Eckhardt, Physica D, 33 (1988) 89.\\ \ \\
32. C. G\'erard, J. Sj\"ostrand, Comm. Math. Phys., 108 (1987) 391.\\ \ \\
33. T. Ericson, Phys. Rev. Lett., 5 (1960) 430.\\ \ \\
34. T. Ericson, Annals of Physics, 23 (1963) 390.\\ \ \\
35. D.M. Brink, R.O. Stephen, Phys. Lett., 5 (1963) 77.\\ \ \\
36. T. Ericson, T. Mayer-Kuckuk, Ann. Rev. Nucl. Sci., 16 (1966) 183.\\ \ \\
37. W.H. Miller, J. Chem. Phys., 55 (1971) 3150.\\ \ \\
38. W.H. Miller, Adv. Chem. Physics, 25 (1974) 66.\\ \ \\
39. Mesoscopic Phenomena in Solids, Editors B.L. Altshuler, P.A. Lee, R.A.
Webb,
North-Holland, 1991.\\ \ \\
40. L.D. Landay, E.M. Lifshitz, Quanyum Mechanics, 2nd ed., Pergamon, New York,
 1965.\\ \ \\
41. C.E. Porter, Statistical Priperties of Spectra, Academic Press, New Your,
 1965.\\ \ \\
42. B.V. Chirikov, Phys. Reports, 52 (1979) 1183.\\ \ \\
43. M.V. Berry, ed., Dynamical Chaos, Cambridge University Press, 1988.\\ \ \\
44. L.E. Reichl, The Transition to Chaos, Springer-Verlag, 1992.\\ \ \\
45. Bibliography on Chaos, complited by Zhang Shu-yu, World Scientific, 1991.\\
 \ \\
46. V.M. Alekseev, Math. USSR, Sbornik, 5 (1968) 73; 6 (1968) 505; 7 (1969)
1.\\
 \ \\
47. J. Moser, Stable and Random Motions in Dynamical Systems, Prinston Univ.
 Press, Princeton, NJ, 1973.\\ \ \\
48. C. Grebogi, E. Ott, J.A. Yorke, Physica D, 7 (1983) 181.\\ \ \\
49. W.H. Miller, J. Chem. Phys., 56 (1972) 38.\\ \ \\
50. M.C. Gutzwiller, J. Math. Phys., 8 (1967) 1979; 10 (1969) 1004; 11 (1970)
 1791; Phys. Rev. Lett., 45 (1980) 150; Physica D 5 (1982) 183; J. Phys. Chem.,
92 (1988)3154.\\ \ \\
51. E. Doron, U. Smilansky, A. Frenkel, Proceedings of
 the International School
 of Physics ``Enrico Fermi", Quantum Chaos,
  North-Holland, 1993, p. 399.\\ \ \\

\end{document}